\documentclass[usenatbib]{mnras}
\defcitealias{PCO15}{Paper I}

\usepackage{graphicx}
\usepackage{amsmath}  
\usepackage{hyperref} 

\newcommand{\amin}{a_\text{min}}
\newcommand{\Tav}{T_\text{av}}
\newcommand{\Teff}{T_\text{eff}}
\newcommand{\Nsph}{N_\text{sph}}
\newcommand{\Mdsk}{M_\text{disc}}
\newcommand{\Mlos}{\dot M_\text{loss}}
\newcommand{\Mrea}{\dot M_\text{reacc}}
\newcommand{\Minj}{\dot M_\text{inj}}
\newcommand{\Ninj}{N_\text{inj}}
\newcommand{\Porb}{P_\text{orb}}
\newcommand{\qr}{q_\text{r}}
\newcommand{\ass}{\alpha_\text{ss}}
\newcommand{\Vorb}{v_\text{orb}}
\newcommand{\Vrot}{v_\text{rot}}
\newcommand{\Sss}{\Sigma_\text{qs}}
\newcommand{\tss}{t_\text{qs}}
\newcommand{\tev}{t_\text{ev}}
\newcommand{\Rt}{R_\text{t}}

\newcommand{\eqt}{Eq.~}
\newcommand{\hdust}{\textsc{Hdust}}
\newcommand{\Ha}{\text{H}\,\alpha}
\newcommand{\Hb}{\text{H}\,\beta}
\newcommand{\Hc}{\text{H}\,\gamma}
\newcommand{\Hd}{\text{H}\,\delta}
\newcommand{\Bg}{\text{Br}\,\gamma}

\newcommand{\tspace}{\hspace{1em}$|$\hspace{1em}}

\author[D.~Panoglou et al.]{
  Despina Panoglou$^1$\thanks{\tt panoglou@on.br},
  Daniel M.~Faes$^2$, Alex~C.~Carciofi$^2$, Atsuo T.~Okazaki$^3$,
  \and~Dietrich Baade$^4$, Thomas Rivinius$^5$,
  Marcelo Borges Fernandes$^1$ \\\\
  $^1$Observat\'orio Nacional, Rua General Jos\'e Cristino 77, S\~ao Crist\'ov\~ao RJ 20921-400, Rio de Janeiro, Brazil\\
  $^2$Instituto de Astronomia, Geof\'isica e Ci\^encias Atmosf\'ericas, Universidade de S\~ao Paulo, Rua do Mat\~ao 1226, SP 05508-900, Brazil\\
  $^3$Faculty of Engineering, Hokkai-Gakuen University, Toyohira-ku, Sapporo, Hokkaido 062-8605, Japan \\
  $^4$European Organisation for Astronomical Research in the Southern Hemisphere, Karl Schwarzschild-Str.~2, 85748 Garching bei \\M\"unchen, Germany\\
  $^5$European Organisation for Astronomical Research in the Southern Hemisphere, Casilla 19001, Santiago 19, Chile
  }

\title[Phase-locked variations in coplanar circular Be binaries]{Be discs in coplanar circular binaries: \\Phase-locked variations of emission lines}

\begin{document} 
\maketitle
\begin{abstract}
The first results of radiative transfer calculations on decretion discs of binary Be stars are presented.
A smoothed particle hydrodynamics code computes the structure of Be discs in coplanar circular binary systems for a range of orbital and disc parameters. The resulting disc configuration consists of two spiral arms, and can be given as input into a Monte Carlo code, which calculates the radiative transfer along the line of sight for various observational coordinates.
Making use of the property of steady disc structure in coplanar circular binaries, observables are computed as functions of the orbital phase.
Orbital-phase series of line profiles are given for selected parameter sets under various viewing angles, to allow comparison with observations. Flat-topped profiles with and without superimposed multiple structures are reproduced, showing, for example, that triple-peaked profiles do not have to be necessarily associated with warped discs and misaligned binaries.
It is demonstrated that binary tidal effects give rise to phase-locked variability of the violet-to-red (V/R) ratio of hydrogen emission lines. The V/R ratio exhibits two maxima per cycle; in certain cases those maxima are equal, leading to a clear new V/R cycle every half orbital period.
This study opens a way in identifying binaries and in constraining the parameters of binary systems that exhibit phase-locked variations induced by tidal interaction with a companion star.
\end{abstract}

\section{Introduction}
The axisymmetry of an ionized keplerian disk manifests itself in time-invariant emission lines with a characteristic double-peak structure. The gravitational field of a companion star is a possible perturber of such symmetry.  It would reveal itself by cyclic perturbations of the violet-to-red flux ratio (V/R) of the two emission components.
Therefore, observations of these so-called V/R variations carry information about potential binarity, the parameters of the companion, and the dynamics and radiative properties of the disc.

Binarity is certainly not a necessity for the production of V/R variations in Be stars, but only an addition to other variability sources \citep{RiBS06}. Be discs in binary systems have been suspected to exhibit tidally-induced phase-locked variations. \cite{KriH75} first studied the binary effect theoretically and in a systematic way, demonstrating that the observational characteristics of Be stars can be a consequence of their binary nature.
Contrary to earlier beliefs \citep{Baad92}, recent observational studies of various binaries reveal variability of different spectroscopic quantities, with a periodicity consistent with the orbital period.
A thorough study of V/R variations in a number of Be binaries and their relation to the orbital period was given by \cite{SORB07}, indicating $\epsilon$\,Cap as a prototype of phase-locked V/R variations.
\cite{SaKu05} suggest that $\kappa$\,Dra is a circular binary with $\Ha$ and $\Hb$ emission locked to the orbital period.
\cite{KuSa08} also report that $\kappa$\,Dra and 4\,Her exhibit V/R variations synchronised with their orbital periods.

\citet[][hereafter \citetalias{PCO15}]{PCO15} studied the Be disc structure in coplanar binary systems (i.e.~in which the disc plane coincides with the orbital plane). A large parameter space was covered, exploring the effects of viscosity, binary mass ratio, orbital period and eccentricity.
The disc is an outward double-armed spiralling flow of gas \citep{OkBa02}.
The azimuthally averaged truncation radius (i.e.~the disc size) increases for higher viscosities, higher orbital periods and lower mass ratios, while due to the accumulation of mass in the region inside the truncation radius, the decline in disc density is shallower than for isolated stars. 

\citetalias{PCO15} also showed that the structure of the disc of Be stars in coplanar binary systems depends on the orbital phase:
Eccentric and/or misaligned (in which there is a tilt between the disc and the orbital plane) binaries are phase-dependent because the structure and the dynamics of the disc depend on the position of the companion in its orbit. In highly eccentric binaries, the disc is almost completely dissolved at the periastron passage (see \citealt{ReFC97}).
In a circular binary with a disc in its asymptotic state, the disk configuration does not change with time, but the way it presents itself to the observer does evolve periodically due to the spiral arms.
The aim of this paper is to demonstrate that spectroscopic variations of Be stars can be attributed to this phase-dependent disc structure.

The outline of this paper is as follows: First (\S\ref{s:fits}), the basic properties of coplanar circular binaries are confirmed, by examination of the azimuthal density structure for systems of various parameters.
Next, the first results of radiative transfer calculations for different parameter sets are presented (\S\ref{s:rt}). Discussion of certain features that were obtained through the simulations, along with some limited comparison to observed stars, is given in \S\ref{s:dis}.
In \S\ref{s:end}, the conclusions are summarised and a basic framework of the variability scheme in Be stars is presented.

\section{Corotating structure at steady state}\label{s:fits}
The effect of the tidal interaction between the two components of a binary system is the truncation of the disc. \cite{OkBa02} showed that truncation does not mean that the disc is really broken at some distance smaller than the orbital separation (see also \citealt{ReFC97}). Rather, beyond the truncation point the density drops much more rapidly with increasing distance from the primary Be star, i.e.~after the truncation point the disc becomes immensely more tenuous.
Under these considerations, it is possible to define the \emph{truncation radius} $\Rt$ (which quantifies the effective disc extent), the \emph{inner-disc surface-density drop-off exponent} $m$, and the \emph{outer-disc density drop-off exponent} $n$ ($n\gg m$).

The gravitational interaction with the companion causes a two-armed spiral wave that breaks the circular symmetry of the disc and makes it time-dependent \citep{Huan72}.
In \citetalias{PCO15} it was shown that when the decretion disc of a binary Be star evolves long enough, it reaches a quasi-steady state (QS) regime, after which the disc structure no longer exhibits cycle-to-cycle variations. Its time dependence can be substituted with a simple orbital phase dependence.
The surface density can be given as a function of position $(r,\phi)$ and time $t$, \mbox{$\Sigma(r,\phi,t>\tss)=\Sss(r,\phi,p)$}, where $\tss$ is the time needed for the system to reach QS, and:
  \begin{equation}
  \Sss(r,\phi,p)
  = A(\phi,p)\frac{(r/\Rt(\phi,p))^{-m(\phi,p)}}
  {1+(r/\Rt(\phi,p))^{n(\phi,p)-m(\phi,p)}},\label{e:Atsne}
  \end{equation}
where $r$ is the radial distance from the central star, $\phi$ is the azimuthal coordinate, $p$ is the orbital phase, $\Rt$ is the truncation radius, $m$ and $n$ are the inner and outer disc exponents, respectively, and $A$ is a parameter related to the base density. The quantities $\Rt,m,n,A$ are numerically fitted for each azimuthal angle $\phi$, and are phase-dependent.
\eqt\eqref{e:Atsne} was originally introduced by \citet[their \eqt14]{OkBa02} for the azimuthally-averaged surface density of the disc. In the above form, the exponent $n$ is slightly redefined so that it really translates to the slope in the outer region of the disc, and the parameters are given as functions of $\phi$ and $p$.

In \citetalias{PCO15} the simulated surface density was fitted to \eqt\eqref{e:Atsne} for each azimuthal direction $\phi$, reaching the conclusion that in coplanar circular binaries the disc structure does not change in shape with time, but only rotates in phase with the secondary.
Mathematically, the azimuthal rotation of the disc structure per orbital cycle can be expressed as
  \begin{equation}\label{e:orbmod}
  x(\phi,p+\Delta p)=x(\phi-\Delta\phi,p) \,\text{with}\,
  \left\{	\begin{array}{l}
  \Delta\phi=\Delta p\cdot360\degr \\
  x\in\{\Rt,m,n,A\}\end{array}.
  \right.\end{equation}

\begin{table*}
\caption{(a) SPH simulation parameters for the binary systems presented in this work: identification number for each simulation (ID; the ID number indicates a single set of simulation parameters throughout the text), orbital period $\Porb$, orbital separation $a$, mass ratio $\qr$, viscosity $\ass$, (constant) number $\Ninj$ of particles injected into the disc base per time step, mass of each particle $\Delta m$, duration of simulation time step $\Delta t$, total evolution time $\tev$.
(b) SPH variables estimated a posteriori: time to reach the steady state $\tss$, number of SPH particles in the end of the simulation $\Nsph$, final number density at the disc base $n_0$, total disc mass $\Mdsk$, stellar mass loss rate $\Mlos$.}
\label{t:pars}
\begin{tabular}{@{}c@{\tspace}ccccc@{ }ccc@{\tspace}ccc@{ }c@{ }c@{}}
\hline
\multicolumn{9}{c|}{(a) SPH input} & \multicolumn{5}{c|}{(b) SPH output}  \\
\hline
ID &$\Porb$ & $a$ & $\qr$ & $\ass$ &$\Ninj$ &$\Delta m$ & $\Delta t$
& $\tev$ & $\tss$ & $\Nsph$ & $n_0$ & $\Mdsk$   & $\Mlos$  \\
\# &(d) &($R_*$) & &&&($10^{-14}M_\odot$)
&(d) & ($\Porb$) & ($\Porb$) & & ($10^{13}$ cm$^{-3}$) &
($10^{-9}M_\odot$) & ($10^{-9}M_\odot$/yr)\\
\hline
\hline
42& 30&17.0&0.08&0.1& 400&3.271&0.048& 93& 84&39412&9.108& 12.920& 1.00 \\
31& 30&17.0&0.08&0.4&2000&0.654&0.048& 51& 42&46158&7.294& 3.029 &0.14 \\
37& 30&17.0&0.08&1.0&4000&0.327&0.048& 25& 21&41089&6.738& 1.344 &0.07 \\
66& 10& 8.2&0.08&0.4& 800&0.545&0.016& 69& 64&39330&7.960& 2.144 &0.31 \\
36& 30&20.8&1.00&0.4&2000&0.654&0.048& 67& 28&46700&7.587& 3.071 &0.54 \\
\hline
\end{tabular}
\end{table*}

In this work the fitting procedure followed in \citetalias{PCO15} was improved by adding radial power-law weights of the form $w(r)=b^r$, with $b=3$.
The basic motivation for this change was that the fitting weights in \citetalias{PCO15} were almost always selected by eye (test-and-correct) for each angle. The new weight function works satisfactorily not only for all angles at a given evolution time in a simulation, but also for the vast majority of simulations (over a wide range of simulation parameters, e.g.~resolution, injection rate, disc viscosity, orbital period, mass ratio, eccentricity).
The new results on the azimuthal structure of the disc and how it is affected by the values of the parameters confirm the conclusions of \citetalias{PCO15}, but they are more accurate and allow for better distinction of the differences between binary systems of different parameters.

\begin{figure}\centering
\includegraphics[scale=.65]{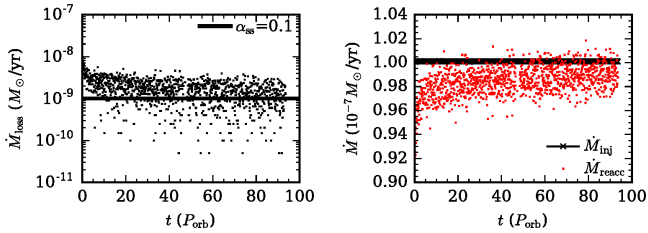}
\caption{\emph{Left}: The mass loss rate $\Mlos$ from the star as a function of time; the horizontal line is the mean stellar mass loss rate after the initial stages, which at QS coincides with the mass loss rate from the disc, so that the total disc mass remains constant. \emph{Right}: The mass reaccretion rate $\Mrea$ (dots) is scattered below the mass injection rate $\Minj$ (constant; horizontal line).
In order to reduce both the number of dots displayed and their scattering, each dot corresponds to the average of 5 consecutive time steps. Both plots are for a system with $\ass=0.1$, $\Porb=30$ d, $\qr=0.08$ (\#42).}\label{f:Mdots}
\end{figure}

Three-dimensional smoothed particle hydrodynamics (SPH) simulations were performed for binary systems in which the Be disc is formed by uniform ejection of matter along the equator \citep{OkBa02}.
The mass injection rate is constant and the same in all simulations, \mbox{$\Minj=10^{-7}M_\odot/$yr}. The mass loss rate $\Mlos$ at QS roughly scales with $\Minj$, but is much smaller (about \mbox{2-3} orders of magnitude in the presented simulations) because most of the injected particles are reaccreted back onto the star (\autoref{f:Mdots}).
The value of $\Minj$ was selected because it ensures a reasonable value for $\Mlos$ (\mbox{$10^{-12}-10^{-9}M_\odot$/yr}; \citealt{Krti14}), as shown in \autoref{t:pars}.

The total mass injected per time step is divided into a fixed number $\Ninj$ of particles of equal mass $\Delta m$ (\autoref{t:pars}). For each SPH simulation, the values of $\Ninj$ and $\Delta m$ were chosen such that an adequate final number of particles at QS (\mbox{$\Nsph\sim4\times10^4$}) is achieved.
The disc is considered isothermal and has a fixed average temperature equal to $\Tav=0.6\Teff$ with $\Teff=19370$~K. The primary Be star has mass $M_*=11.2M_\odot$ and radius $R_*=5.5R_\odot$ (typical values for a B2 star). Unless stated otherwise, the orbital period is $\Porb=30$ d, the secondary-to-primary mass ratio is $\qr=M_2/M_1=0.08$, and the \citeauthor{ShSu73} viscosity parameter is $\ass=0.4$.

\begin{figure}\centering
\includegraphics[scale=.63]{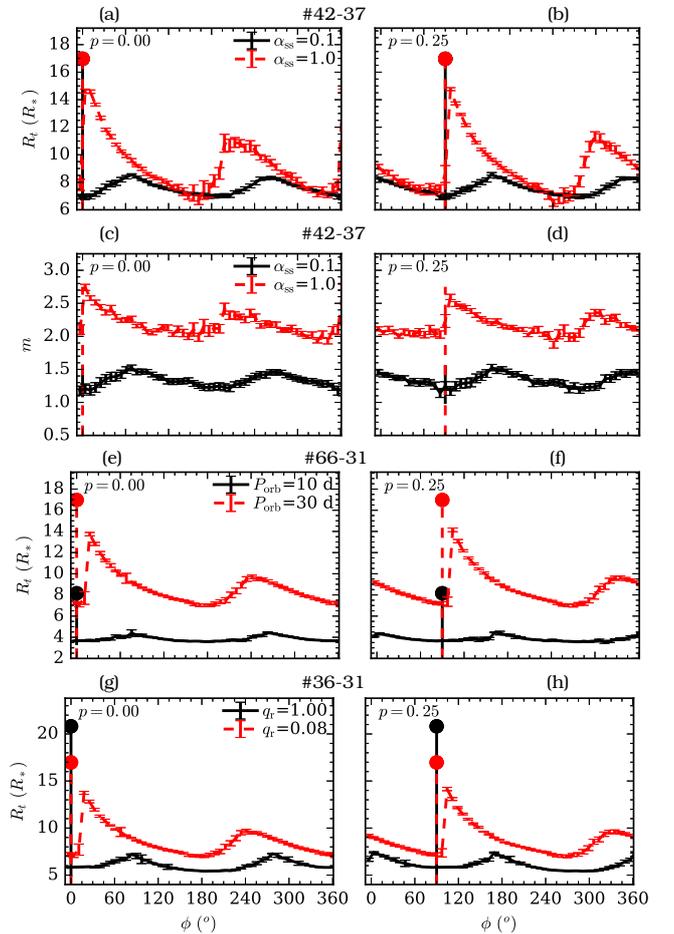}
\caption{Representative plots of parameters from \eqt\eqref{e:Atsne} as functions of the azimuthal angle, for simulations of different parameter sets (each row of panels shows the results for two simulation IDs, as labelled on the top of the row, whose parameters are given in \autoref{t:pars}). The error bars denote the errors in the fitting procedure, in which the surface density values were folded for 5 cycles at QS.
Each panel corresponds to a single phase (left: $p=0$; right: $p=0.25$), with the vertical line showing the direction to the secondary and the filled circle (in case of $\Rt$ plots) indicating its position (i.e.~its distance from the Be star).}\label{f:Rta}
\end{figure}

The truncation radius $\Rt$ (Figures \ref{f:Rta}a-b, e-f, g-h) reaches a local maximum twice per cycle. The global maximum follows the direction of the secondary with some time lag, and the other local maximum is antidiametric to the first one. The two local maxima of $\Rt$ trace the two spiral arms of the disc. The spiral arm that follows the secondary is generally different (more prominent) to its antidiametric one.
The same azimuthally-shifted dependence on the orbital phase is clear for the other parameters of \eqt\eqref{e:Atsne}.

\begin{itemize}
\item[]It is highlighted that:
\item For higher values of viscosity the spiral arms are less tightly wound, i.e.~larger variation of $\Rt$ (\autoref{f:Rta}a) and higher $m$ values (\autoref{f:Rta}c). The disc extends to larger distances from the star, and the time lag of the global maximum of $R_t$ is minimal.
\item \autoref{f:Rta}e confirms that a smaller orbital separation results in a smaller azimuthal modulation of the disc extent. In very close binaries the disc is almost circular (\autoref{f:Rta}e, $\Porb=10$~d).
\item A smaller secondary-to-primary mass ratio (less massive companion) results in disc truncation further away from the star, as shown in \autoref{f:Rta}g, although the orbital separation is smaller (\autoref{t:pars}). This can be explained as follows: Lower mass ratio means weaker gravitational force from the companion, and the disc is saturated less efficiently. The contrary happens for higher mass ratios, especially in the outer parts of the disc.
Therefore truncation occurs at a longer distance from the primary star in systems of lower mass ratios. A smaller mass ratio also causes less closely wrapped spiral arms, but a smaller azimuthal lag with respect to the position of the secondary.
\item The simulations confirm the correlation between the time-scale of the disc variability and the orbital period found by \cite{ReNe05} in Be/X-ray binaries.
\end{itemize}

Note that, when the disc size is smaller (lower viscosity, closer binary, higher mass ratio), the azimuthal disc structure exhibits a variability frequency almost twice the orbital frequency, as the two spiral arms have a similar structure with almost equal high peaks.
In other words, the two maxima in the functions of the parameters of \eqt\eqref{e:Atsne} have equal values. This indicates a higher similarity in the density structure of the two spiral arms. It is expected that this (at least partially) is reflected also on the observables, and it will be evinced in the next section for the viscosity.

\section{Radiative transfer calculations}\label{s:rt}
At any time of evolution calculated with the SPH code, the density structure of the disc can be used as input for the three-dimensional non-local thermodynamic equilibrium Monte Carlo radiative transfer code {\hdust} \citep{CaBj06}. {\hdust} first computes the temperature, ionisation and excitation structure of the gas, and then calculates the emergent spectrum for chosen directions.

\begin{figure}\centering
\includegraphics[clip,trim=25mm 11cm 88mm 16mm,scale=.88]{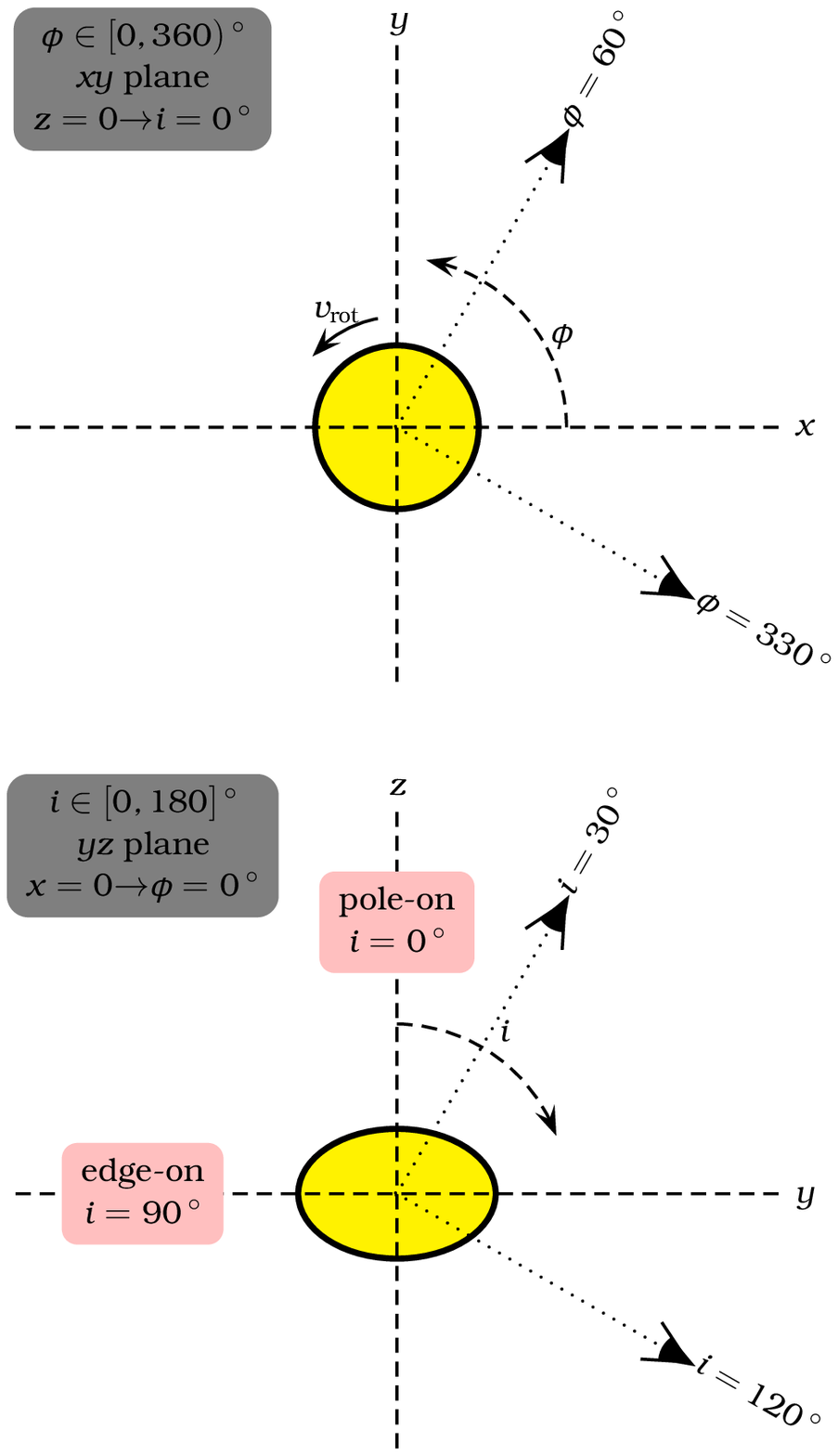}
\caption{The geometry of the system. \emph{Top}: The equatorial plane ($xy$), on which the azimuthal angle $\phi$ is measured. Since the binary systems presented here are all coplanar, the equatorial plane coincides with the disc and orbital planes. \emph{Bottom}: A plane that passes through the rotational axis $z$ of the Be star at an arbitrary angle $\phi$ (i.e.~plane perpendicular to the equatorial plane).
Over any such plane the inclination (viewing) angle is measured.}\label{f:geom}
\end{figure}

\begin{figure}\centering
\includegraphics[scale=.65]{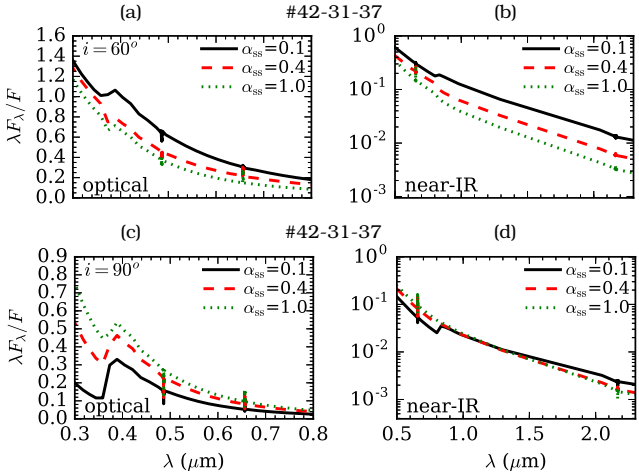}
\caption{The emergent spectrum in the visible (left) and near-IR (right) bands at phase $p=0$ for a disc with different values of viscosity (labelled on the right panels), as seen at $i=60$ (top) and $90\degr$ (bottom).}\label{f:obs60}
\end{figure}

\begin{figure}\centering
\includegraphics[clip,trim=0mm 2mm 0mm 0mm,scale=.72]{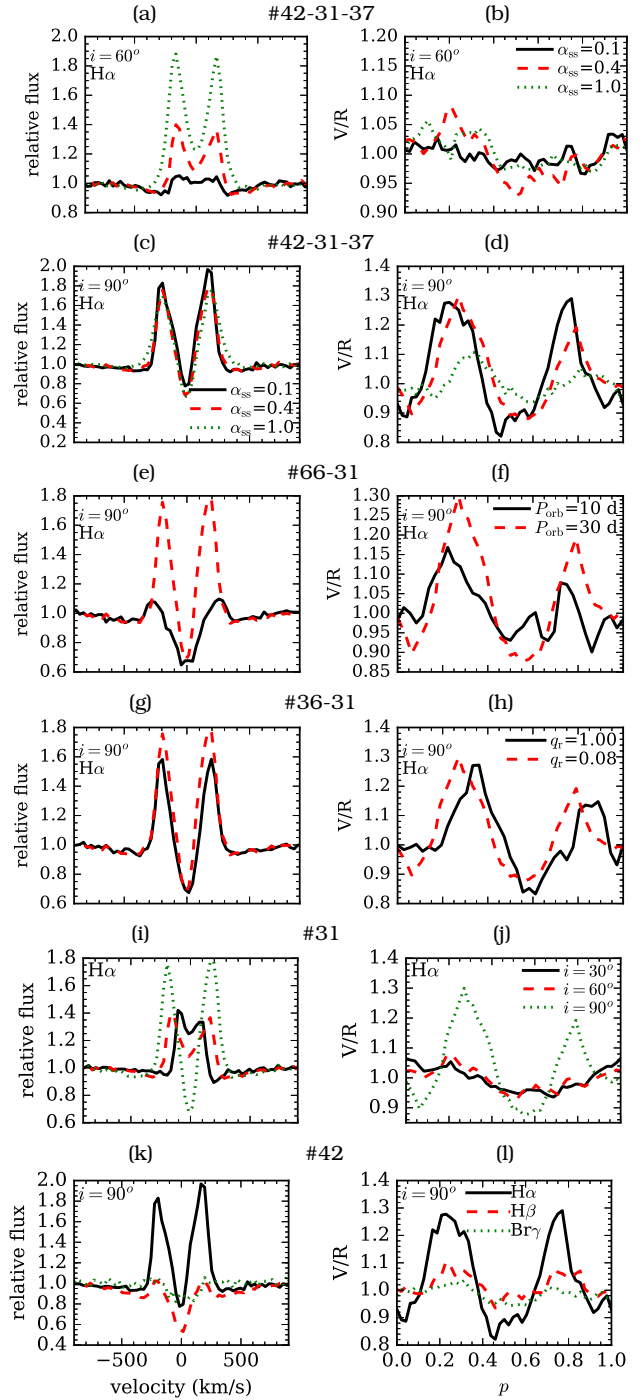}
\caption{Several emission-line profiles at phase $p=0$ (left) and the V/R peak ratio as a function of $p$ (right) for systems of various parameters and observational inclinations (as labelled), after the disc has reached QS. The labels are valid for both panels of the same row.}\label{f:obsHao}
\end{figure}

For an observer seeing the Be star at inclination $i$ and azimuthal angle $\phi$ (the observer's coordinates are clarified in \autoref{f:geom}), {\hdust} can compute the radiation flux for a range of wavelengths, as well as the line profiles for selected hydrogen transitions.
Given the orbital modulation (\eqt\ref{e:orbmod}) of the Be disc density at QS, the azimuthal angle of the observer can be transformed to a phase difference. Thus, running {\hdust} simulations for observers at constant $i$ but different values of $\phi$, it is possible to compute functions of the orbital phase for various observables.

For the {\hdust} simulations, the primary Be star was considered oblate, with gravity darkening parameter \mbox{$\beta=0.19$} \citep{EslR11} and rotational-to-orbital velocity ratio $\Vrot/\Vorb=0.7$. The rest of the parameters for an oblate star are calculated as explained by \cite{Faes15}: equatorial and polar radii 5.5 and $4.5R_\odot$, equatorial and polar temperatures 18605 and 22670 K, respectively.

The radiative flux is higher for lower viscosities (\autoref{f:obs60}), as lower viscosity discs are denser (\autoref{t:pars}).
This results in more light being absorbed at short wavelengths (\mbox{$\lambda<0.365~\micron$}, Balmer jump) and re-emitted at longer wavelengths, producing a respective flux excess in the visible and infrared (IR).
An exception holds at $i\simeq90\degr$: for an edge-on star (\autoref{f:obs60}c), the denser low-viscosity disc blocks a considerable portion of the stellar flux in the visible. Since a denser disc is smaller and cooler, it causes a higher flux excess only at longer wavelengths (far-IR).
For lower values of $\ass$ at $i<90\degr$, the higher continuum level makes the relative emission intensity in $\Ha$ lower, as shown in \autoref{f:obsHao}a.

The V/R ratio is defined as the ratio of the violet and red peak heights (relative to continuum) in a double-peaked line profile:
  \begin{equation}\text{V/R}=\frac{F_\text{v}/F_\text{c}}
                                  {F_\text{r}/F_\text{c}}\end{equation}
where $F_\text{c}$ is the continuum flux, $F_\text{v}$ is the peak emission of the V component, and $F_\text{r}$ is the peak emission of the R component.
Symmetric profiles (V/R=1) indicate axisymmetric discs, while asymmetries arise from non-symmetric disc configurations and radial motions. The $\Ha$ V/R variability amplitude increases with decreasing disc viscosity (\autoref{f:obsHao}d).
Since different parts of the disc, in general, have different velocities along the line of sight (LoS), it is expected that asymmetries in a spectrum are a combined result of both the disc structure and the projected velocities that are observed.
Therefore, for inclinations $i<90\degr$, where the V/R variability amplitude is so small (at least for the range of parameter values explored in this work), it is difficult to constrain the system properties from this observable alone (\autoref{f:obsHao}b).

\begin{figure}\centering
\includegraphics[scale=.65]{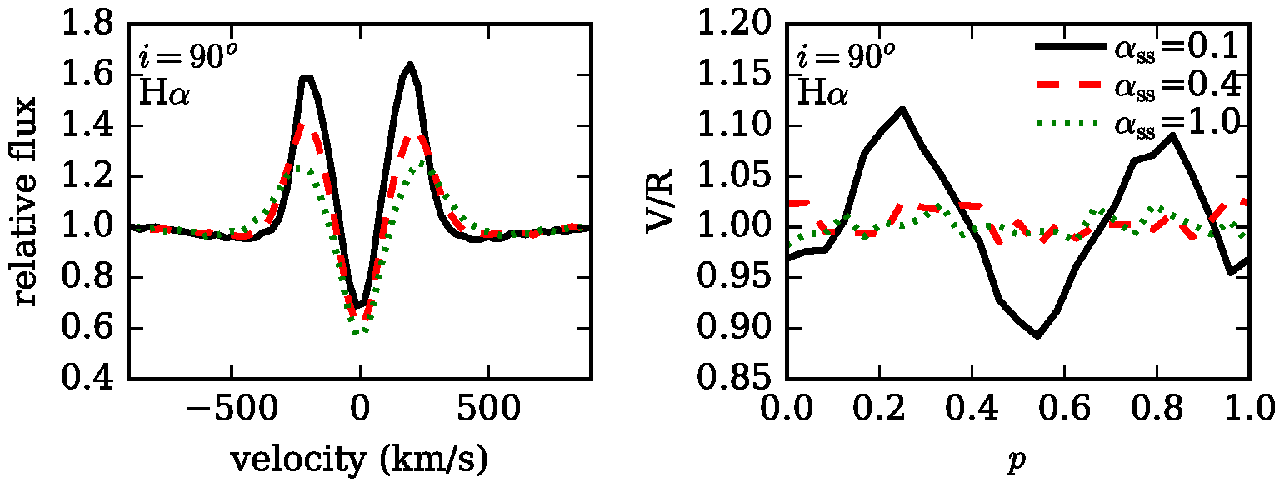}
\caption{Same as \autoref{f:obsHao}b at $i=90\degr$ with the density scaled to $n_0=10^{13}$ cm$^{-3}$.}\label{f:obsHan0}
\end{figure}

In order to exclude the possibility that the aforementioned effect of viscosity (i.e.~V/R variability decreasing with increasing viscosity) is a consequence of the different densities in SPH simulations, new radiative transfer calculations were performed. The original density structures from the three systems of different $\ass$ were used, but all three of them were scaled as to have the same density $n_0$ at the disc base (inner disc layer).
In this way the continuum flux is overall the same and whatever difference is seen between the relative fluxes in $\Ha$ is due solely to different truncation radii and disc density slopes, as extracted from SPH simulations with different values of disc viscosity. The results are shown in \autoref{f:obsHan0}, affirming that the V/R variability vanishes for high values of $\ass$ even if the base density is fixed.

Figures \ref{f:obsHao}i-j demonstrate the results of radiative transfer calculations for a system with disc viscosity $\ass=0.4$, as seen from different inclination angles.
This intermediate value of viscosity was chosen so that both the difference in the line profiles can be seen (lower values of $\ass$ do not have a double-peaked profile even at $i$ as large as $60\degr$ in the spatial resolution of the SPH simulations; \autoref{f:obsHao}a) and the V/R variability amplitude is sufficiently large at $i=90\degr$ (it is small at higher $\ass$; \autoref{f:obsHao}d).
The results for the emission at different inclinations in general agree with the sample spectra given in figure~1 of \cite{RiCa13} for the line emissions of Be discs.
The $i=90\degr$ curve of \autoref{f:obsHao}i confirms that the line profile of an edge-on star is that of a shell star (i.e.~line profiles having sharp absorption cores below the continuum level; \citealt{HaHu96}).

The V/R ratio in the low-viscosity system has two equal local maxima along the orbital period (\autoref{f:obsHao}d), just as $\Rt$ and $m$ in \autoref{f:Rta}a-d. Hence, if the V/R variation is attributed solely to binary interaction and successive maxima of V/R are equal, its variability period should be considered equal to half the orbital period.
Were it verified that an observed star is a binary and the V/R variability period is equal to half the orbital period, this would hint at higher structural similarity between the two spiral arms.

\begin{figure}\centering
\includegraphics[clip,trim=0mm 0mm 0mm 0mm]{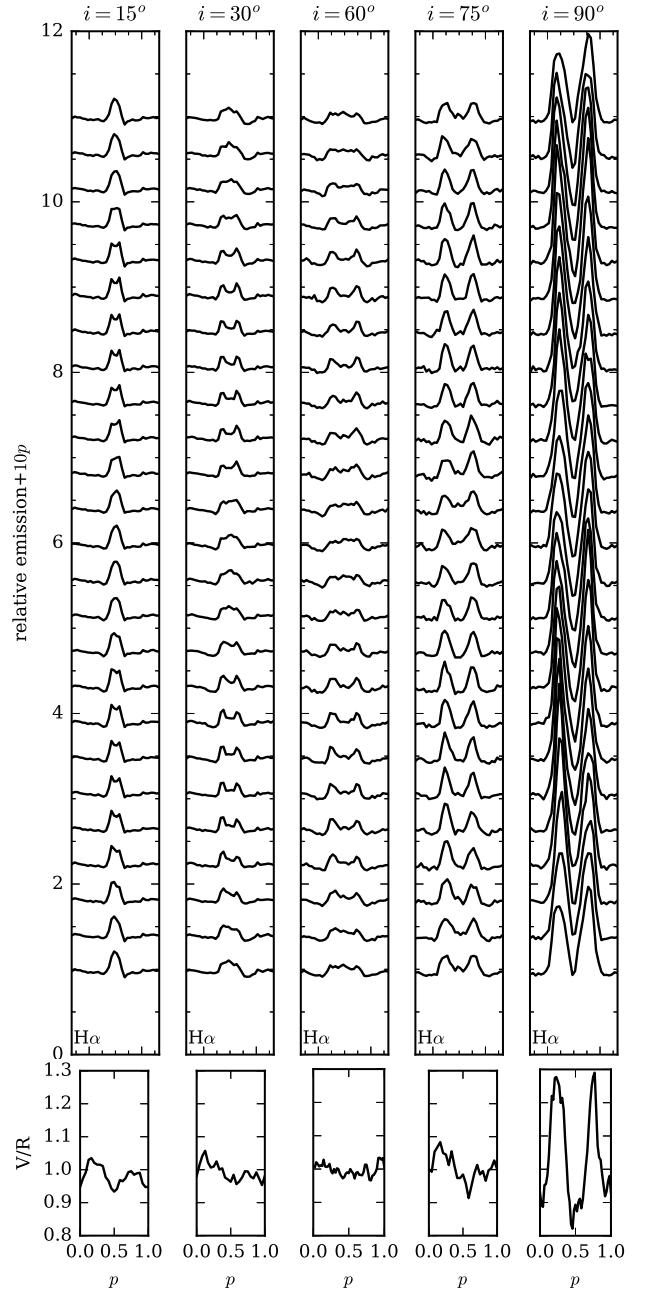}
\caption{\emph{Top:} $\Ha$ line profiles along an orbital cycle from various inclinations (\#42). Continuum levels are given by $10p$, with the orbital phase \mbox{$p\in[0,1]$}. The continuum flux is normalised to unity. Vertical offsets correspond to 4\% in orbital phase. The inclinations shown were chosen based on higher differences between adjacent angles. \emph{Bottom:} The V/R ratio for each inclination angle.}
\label{f:obsds}
\end{figure}

\autoref{f:obsHao}j shows that the V/R orbital variability is more pronounced at higher inclination angles, thus corroborating the suggestion of \cite{PF16} with respect to the importance of inclination for the observables.
This is also verified by \autoref{f:obsds}, which depicts that low-viscosity discs seen edge-on show strong $\Ha$ line emission and V/R variability, while at $i=60\degr$ the profiles are almost flat at any orbital phase.
Note that the $i=60$ and $90\degr$ panels of \autoref{f:obsds} correspond to the same simulation represented by solid curves in Figures~\ref{f:obsHao}a-b and \ref{f:obsHao}c-d, respectively.

\begin{figure}\centering
\includegraphics[clip,trim=0mm 0mm 1mm 0mm]{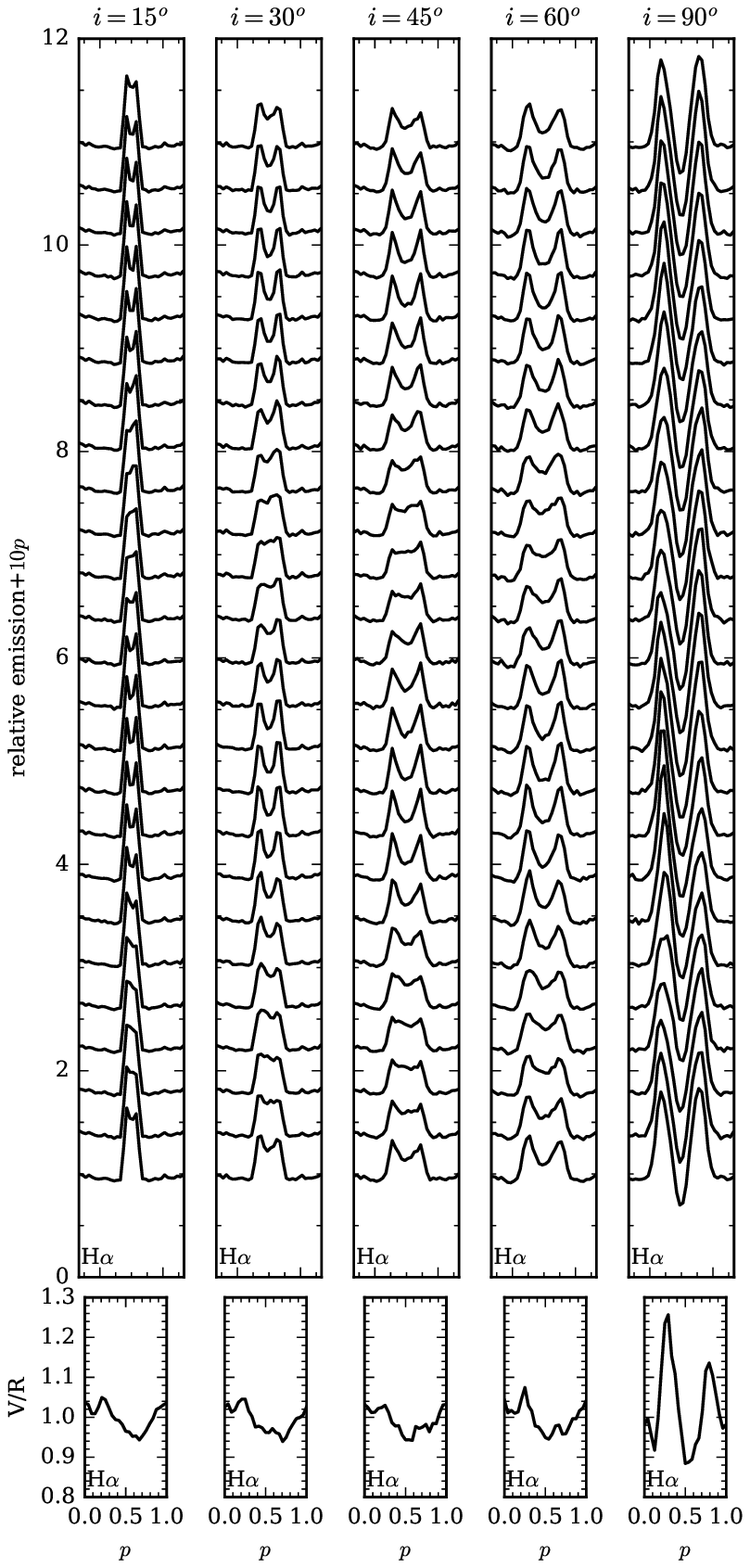}
\caption{Same as \autoref{f:obsds} but for a disc with $\ass=0.4$ (\#31).}
\label{f:obsds.4}
\end{figure}

\begin{figure}\centering
\includegraphics[clip,trim=1mm 0mm 1mm 0mm]{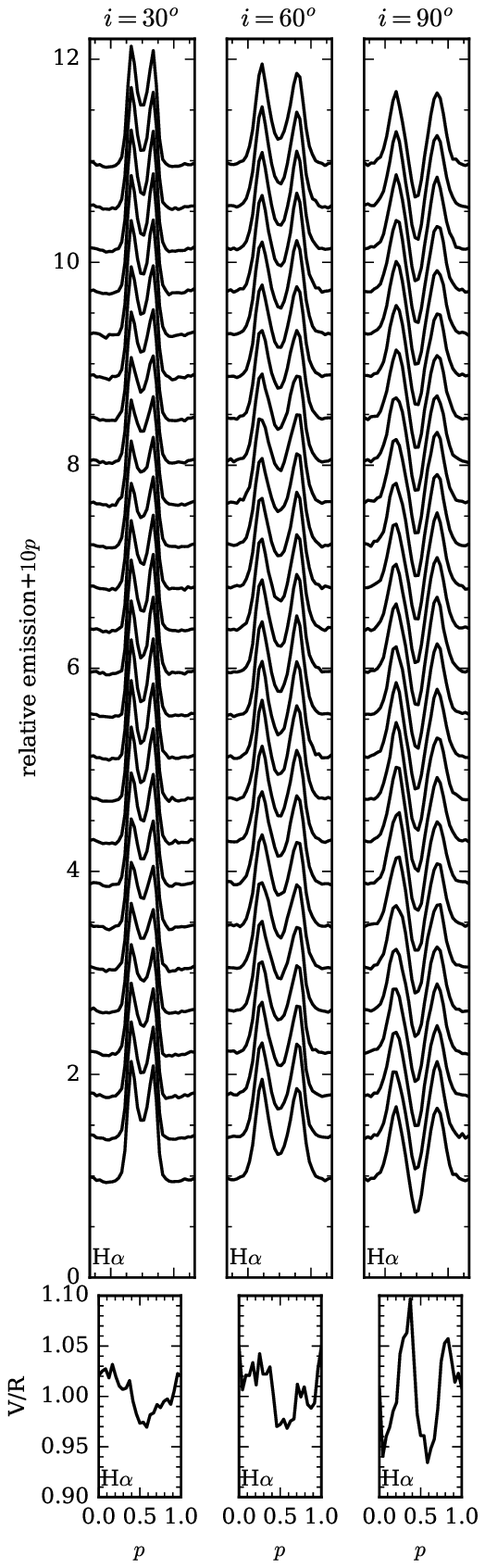}
\caption{Same as \autoref{f:obsds} but for a disc with $\ass=1$ (\#37).}
\label{f:obsds1.}
\end{figure}

For low-to-intermediate values of viscosity, the shape of the line profiles changes both as a function of the orbital phase and with the inclination angle (\autoref{f:obsds.4}; the last two panels of this figure at $i=60,90\degr$ correspond to the dashed curves of Figures~\mbox{\ref{f:obsHao}a-d}, respectively). Some flat-topped profiles appear in $i\in[30,45]\degr$.
For high disc viscosities a simple inspection of the line profiles as they change with the orbital phase (\autoref{f:obsds1.}; the panels in the middle and right column correspond to the dotted curves in Figures~\mbox{\ref{f:obsHao}a-d}, respectively) confirm to a further extent that, for any given viewing angle, there is little or no variability, just as the V/R curves show for the high-viscosity system (\autoref{f:obsHao}d).
This is an indication that, although the disc is highly non-axisymmetric for high viscosities (e.g.~disc size in \autoref{f:Rta}a), the inner emitting region is hardly affected by the tidal interaction with the secondary.
These and other observational characteristics will be further discussed in the next section.

\begin{figure}\centering
\includegraphics[clip,trim=0mm 0cm 0cm 0cm]{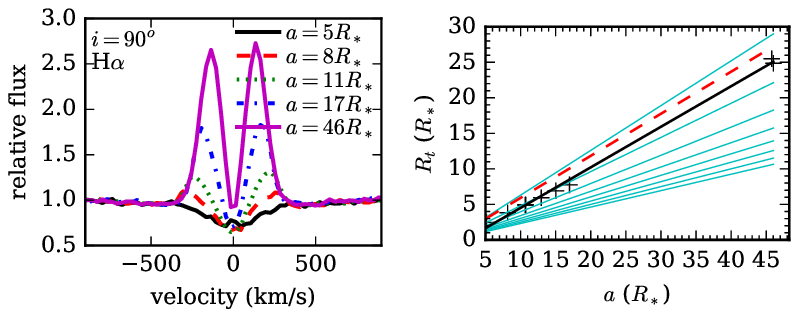}
\caption{\emph{Left}: $\Ha$ profile at phase $p=0$ for systems with various values of the orbital separation (corresponding to $\Porb=5,10,15,30,133$ d, from bottom to top), as seen at inclination $i=90\degr$ ($\ass=0.4$, $\qr=0.08$).
\emph{Right}: The azimuthally-averaged truncation radius as a function of the orbital separation (crosses); the dashed line is the Roche lobe radius of the Be star; the solid dark line is a linear fit (\eqt\ref{e:rta}) of the truncation radius; the series of thinner lines are the 2:1, 3:1, $\ldots$, 10:1 resonance radii, from top to bottom, respectively.}\label{f:rtER}
\end{figure}

In closer binaries the shell features increase and the V/R variability amplitude decreases (\autoref{f:obsHao}e,f). The lower the orbital period/separation, the weaker the line emission, consistent with the fact that the disc is smaller.
The latter is confirmed with the help of \autoref{f:rtER}: The left panel shows the emission for various orbital separations. The right panel depicts the disc size as a function of separation. This figure is a reproduction of the top right panel of figure 12 from \citetalias{PCO15}, with the truncation radius fitted with the new procedure described in \S\ref{s:fits}.
The azimuthally-averaged truncation radius now coincides almost perfectly with the 3:1 resonance radius, and can be expressed as a linear function of the orbital separation $a$:
  \begin{equation}\Rt= 0.57 a -1.20\quad,\quad(\ass=0.4,\qr=0.08)
  \label{e:rta}
  \end{equation}
\par\eqt\eqref{e:rta} provides a lower bound for a star to be able to form a disc: It is defined by the limiting value of disc truncation ($\lim_{\Rt\rightarrow R_*}\Rt$), implying a minimal orbital separation $\amin\simeq3.8R_*$ (minimum $\Porb=3.2$ d).
No disc can be formed in circular binaries closer than $\amin$, whose value probably depends on the other system parameters, as well.
This explains the absence of Be stars in very close binary orbits. In the compilation of \cite{Gies00}, only 4 out of 40 Be binaries belong to the shortest period group, $\Porb\in(4,6)$~d, and they are not even confirmed binaries.
\autoref{f:rtER} confirms the suggestion of \cite{ReFC97} that there is a positive correlation between $\Ha$ emission strength, orbital period and Be disc size.

For higher secondary-to-primary mass ratios, the stronger disc truncation (\autoref{f:Rta}d) makes the emission peak heights slightly decrease (\autoref{f:obsHao}g), but the V/R variability amplitude remains almost the same. A higher mass of the secondary star also causes a larger phase difference of the V/R ratio with respect to the position of the secondary, consistent with the time shift of the truncation radius maxima.

\section{Discussion}\label{s:dis}
In this section, comparisons between the spectra of single and binary Be stars, as well as steady and non-steady Be discs are given. Additionally, some peculiar features that emerged in the emission line profiles of binary Be discs (\S\ref{s:rt}) are investigated, emphasising how they can serve as a first hint at binarity.

\begin{figure}\centering
\includegraphics{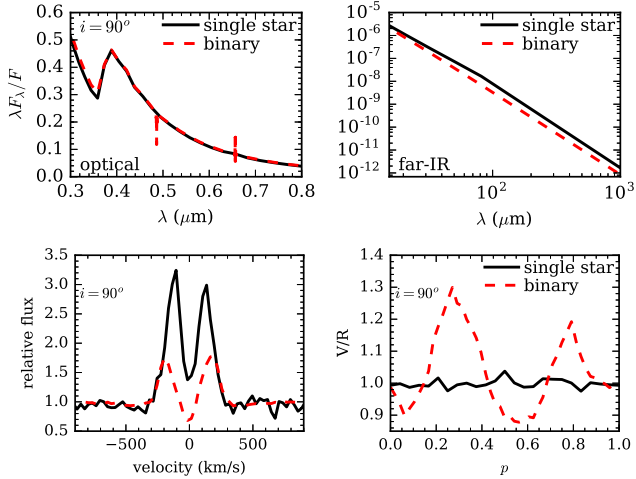}
\caption{The emergent flux at optical (top left) and IR (top right) wavelengths, the $\Ha$ profile (bottom left) and the $\Ha$ V/R ratio (bottom right) for a single and a binary Be star (\#31).}\label{f:obsSB}
\end{figure}

\subsection{Single versus binary stars}
\autoref{f:obsSB} allows to compare the results of radiation transfer calculations for single and binary Be stars. For wavelengths up to the near-IR, the continuum flux is the same in single and binary stars, while in the far-IR the emission for both is a little higher for single stars. The central depression below the continuum level in Be discs in binary systems is due to lack of emission from the outer disc (since it is truncated), where the orbital velocities are low.
No V/R variability is detected in single stars: In the absence of variability sources the disc and its emission are azimuthally symmetric.

\subsection{Non-steady discs}
Until now, only discs at QS have been studied. But, along with binary phase-dependent effects, other sources of variability might be also at play. The figures throughout this text should therefore be used as a reference only, keeping in mind that additional variability sources will add to or diminish the variations found in the model spectra.
In this continuous interplay between the different mechanisms that may cause variations in the line profiles of Be stars, QS is an idealised scenario.

\begin{figure}\centering
\includegraphics[scale=.65]{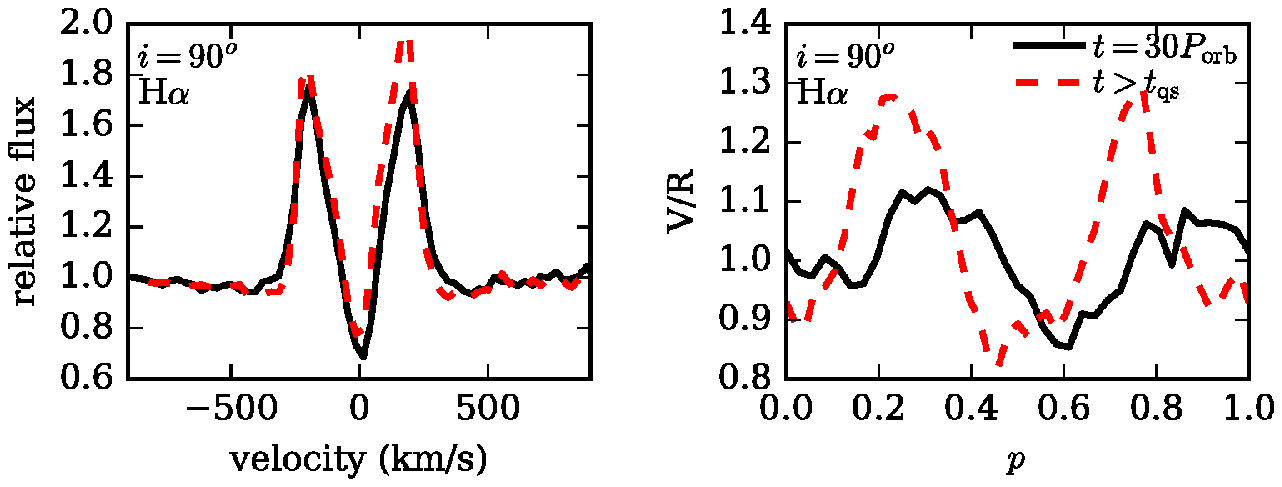}
\caption{An $\Ha$ line profile at $p=0$ (left) and the V/R ratio (right), for the same binary system (\#42) but at two different epochs: one at QS ($t>\tss=84\Porb\simeq7$ yr) and one at \mbox{$t=30\Porb\sim2.5$~yr}.}\label{f:obsNqss}
\end{figure}

The majority of observed systems are probably out of QS. For this reason, a non-QS system was also studied (\autoref{f:obsNqss}). At $\sim1/3$ of the time necessary to reach QS, the $\Ha$ profile peaks are similar to QS, but the V/R variability is relatively weak. This is a direct effect of the early evolution of a disc in a binary system.
The disc is still evolving towards its final spiralled QS structure and the spiral arms are not prominent enough yet. Therefore the $\Ha$ emitting region is almost axisymmetric with no significant azimuthal dependence.

Even pronounced V/R variations in confirmed Be binaries can be not at all synchronised with the orbital period. A disc under the tidal effect of a companion star needs some time to become steady and maximise its V/R variability amplitude. The time to QS is delayed by other mechanisms that disturb its evolution. The most important factor in this respect is probably the generally variable mass ejection rate from the surface of the star.
As a result, a Be disc in a binary system will reach QS later than theoretically computed (\autoref{t:pars}).
When it does reach QS, at any time it can be perturbed and then start again to relax towards a steady state.

That might explain why, in many cases of confirmed binaries, the principal variability cycles have nothing to do with the binary period. For instance, the binary Be star $\zeta$\,Tau exhibited a V/R variability of $\sim1400$~d for more than a decade \citep{SRCl09}, but its orbital period is $\sim133$~d.

\begin{figure*}\centering
\includegraphics[scale=.65]{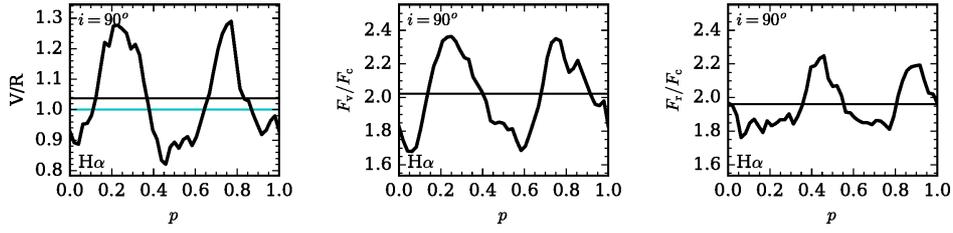}
\caption{The V/R ratio (left) and the V (centre) and R (right) components disentangled for the $\Ha$ profile (\#42). The dark horizontal lines depict the average values along the orbital cycle.}\label{f:VR}
\end{figure*}

\begin{figure}\centering
\includegraphics[clip,trim=5mm 5mm 5mm 12mm,scale=.65]{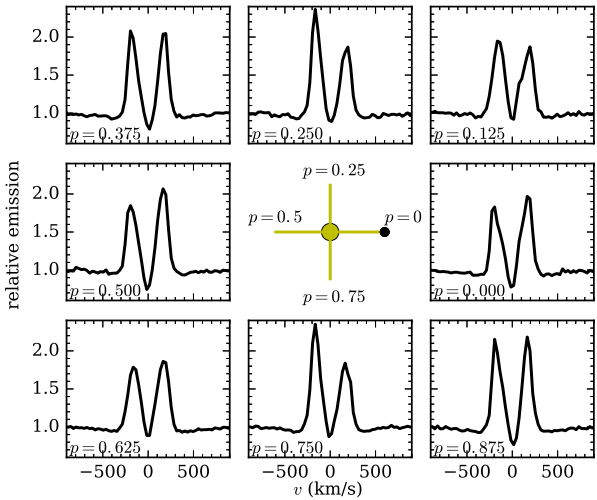}
\caption{$\Ha$ profiles (\#42) along an orbital cycle, spaced by \mbox{$\Delta p=0.125$}. A pole-on view of the Be star is shown at the centre of the figure. The observer looks at the system from an edge-on orientation. At $p=0$ the companion (black circle) is between the Be star and the observer. As the companion orbits the Be star, the profile as received by the observer in different orbital phases is plotted.
At phase $p$ of each profile, the azimuthal angle between the LoS and the line that connects the two stars is given by $\phi=p\cdot360\degr$.}\label{f:perf}
\end{figure}

\subsection{Line-profile features}
\subsubsection{Overall asymmetries}
The gas in the spiral arms (which is denser than in the inter-arm regions) in general follows the direction of the companion's orbit. If each arm were to be observed alone, it would be seen as blue-shifted (V$>$R) for some phase range in the orbital cycle and red-shifted (V$<$R) for the rest.
If one of the arms is more prominent, then both the V and R emissions will be higher for this arm. This is always true in a two-armed binary-induced structure: the spiral arm that directly follows the companion star is denser than its antidiametric one.

The observer cannot see each one of the spiral arms isolated, but rather sees the disc as a whole with two (one dense and one more tenuous) spiral arms, together with the inter-arm gas. In the region between the two spiral arms, the particles are scattered in terms of velocity values and directions.
Every half orbital period, a different spiral arm is directly seen by the observer. In the time between the direct observation of either one of the spiral arms, mostly the scattered gas particles are seen.

A more detailed description can be read off from Figures \ref{f:VR}-\ref{f:perf}, by following therein the LoS during the companion's orbital path for an edge-on star. At $p=0$, the companion star is at lower conjunction and at $p=0.5$ at upper conjunction.
At $p=0$, the more massive and dense spiral arm is receding from the observer (red-shifted regime), and at $p=0.2$ its tail and the other arm are seen approaching so that its emission is at positive LoS velocities (V/R$>$1, blue-shifted regime). At $p\simeq0.5$ there is a new minimum, related to the latter (more tenuous) arm now receding.

At a first glance, it seems surprising that the two emission peaks are not equal even when averaged over a full orbit (\autoref{f:VR}). This is not associated to the relative rotation of the disc gas with respect to the rotation of the companion.
As explained in \citetalias{PCO15}, the disc in circular retrograde systems is axisymmetric and simulations confirm that there is no V/R variability in such systems, just as in simple unperturbed keplerian rotation of single stars.
Therefore the observational variability is restricted to prograde orbits.

So what causes the net orbit-averaged asymmetry of the double-peaked profile? The answer lies in the relative rotation of the system with respect to the observer. If the system were rotating in the direction opposite to the one shown in \autoref{f:geom}, i.e.~if the stellar rotational axis rotated by $180\degr$, the V/R curve would be flipped about both abscissa and ordinate.
In other words, the preferential direction of the asymmetry in the V/R curve encodes the information whether the disc is seen from below or from above. This can be observationally tested through interferometry with phase information.

\subsubsection{Flat-topped line profiles}\label{s:ftp}
In certain phase ranges, the inter-arm gas together with the two arms (one red- and one blue-shifted) produce the total emission reaching the observer as flat-topped profiles.
Figures \ref{f:obsds} and \ref{f:obsds.4} show that flat-topped profiles are common in binary Be stars with low/intermediate disc viscosities and seen at intermediate viewing angles.
The reason for this behaviour is probably connected to the observed differential velocities and the spiralled structure of the disc.

Such profiles occur twice per orbital period. This remark becomes more obvious in \autoref{f:obsds.4}, where the emission peaks in non-flat-topped profiles are higher and therefore the variation with the orbital phase is more clear.
A closer look at the $i=30\degr$ panel of \autoref{f:obsds.4} allows to recognize that the 2nd-4th profiles from the bottom of the plot, which correspond to $p\in(0.04,0.12)$, are flat-topped and exhibiting a slope towards the violet side. The 13th-15th profiles, which correspond to $p\in(0.48,0.6)$, are also flat-topped and leaning to the red side.
The mean values of $p$ in those phase ranges ($p=0.08, 0.54$) indicate that this transition occurs nearly every half orbital period.

\begin{figure}\centering
\includegraphics[clip,trim=0cm 7cm 9cm 0cm,scale=.7]{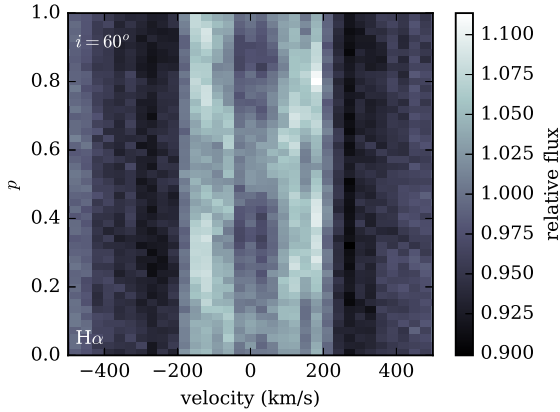}
\caption{Dynamical spectra for $\Ha$ (\#42), seen with inclination $i=60\degr$.}
\label{f:ds.1}
\end{figure}

Flat-topped profiles have been observed in the past, e.g.~the $\Ha$ profiles given for HR\,2142 in figure~23 of \cite{HaHu96}. HR\,2142 is a binary system \citep{Pete83} consisting of a Be star and a subdwarf O star, with period \mbox{$\Porb\simeq81$~d}, mass ratio $\qr\simeq0.07$, and presumably zero eccentricity \citep{PeWa16}.
All binary parameters are therefore similar to the $\ass=0.1$ simulation (\#42), except for the orbital period. The dynamical spectrum of this system at inclination angle $i=60\degr$ (\autoref{f:ds.1}) matches well its observational equivalent \citep[][bottom panel of their figure~4]{PeWa16}. \cite{PeWa16} estimate the inclination angle within the range $65-85\degr$.

The profiles of \cite{HaHu96} for HR\,2142 are similar in shape to the profiles of \autoref{f:obsds} (at $i=60\degr$), but not in intensity (emission height). As shown in \autoref{f:rtER}, the emission height scales with the orbital separation. Therefore, higher emission profiles are qualitatively expected, as the orbital period of HR\,2142 is much longer than 30 d (\#42).
Furthermore, the line emission changes with the inclination angle (\autoref{f:obsHao}i), hence an inclination slightly different to $60\degr$ might increase the $\Ha$ line emission but roughly preserving the overall shape. This could enhance the assumption of higher inclination by \cite{PeWa16}.

The consecutive blueward and redward transitions follow the disc rotation and essentially are LoS effects, as also \cite{SORB07} have stated.
V/R does not take on values much lower than unity because, at the same time that the dense spiral arm comes closer to the observer, the tenuous spiral arm recedes. Whatever moves away contributes to the red-shifted emission component; whatever approaches contributes to the blue-shifted component. Oscillating flat-topped profiles occur during a narrow phase interval in which the number of approaching and receding particles is about balanced.

In summary, cyclically recurring flat-topped emission lines that change from a blueward to a redward declining slope may be indicative of the presence of a companion star.
One such example is 25\,Ori (Baade et al., in prep.), which has never in the past been a suspected binary. If the variability of 25\,Ori is indeed induced by a companion, from the profiles in Baade et al.~the orbital period can be estimated to about 6~yr and the minimum separation with the companion 7.2 AU.

\begin{figure}\centering
\includegraphics[clip,trim=1mm 0mm 1mm 0mm]{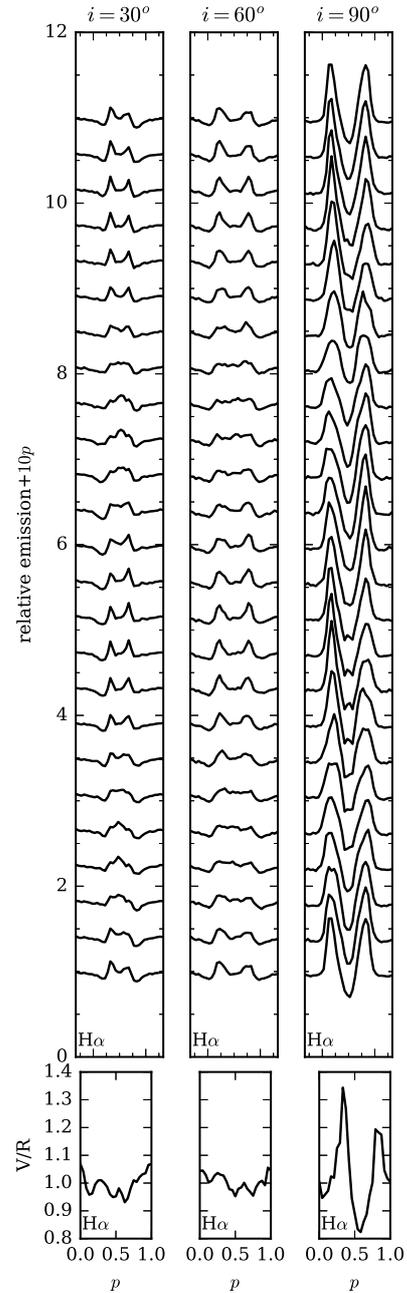}
\caption{Same as \autoref{f:obsds} for a binary system of mass ratio $\qr=1$ (\#36), seen from various inclinations.}
\label{f:obsdsq1}
\end{figure}

\subsubsection{Inflection points}
As shown in \autoref{f:obsHao}h, the V/R variability amplitude stays roughly the same for a system with more than 10 times larger mass ratio. Line profiles for the $\qr=1$ simulation (\#36) were also computed for different orbital phases (\autoref{f:obsdsq1}).
Flat-topped profiles appear twice per period at $i=30$ and $i=60\degr$. At $i=30\degr$, the shape changes from typical, in high inclinations, (non-symmetric) double-peaked profiles (bottom profile) to profiles with three peaks (4th-5th and 16th-17th profiles), the middle one of which is higher than the other two.
The phase range of transitioning between double-peaked and triple-peaked profiles is characterised by flat-topped profiles (3rd, 6th and 15th, 18th, respectively).

According to its mathematical definition, an \emph{inflection point} is where the curvature of a continuous function changes direction, i.e.~where its second derivative changes sign. Changes of the curvature direction occur in flat-topped and triple-peaked profiles, producing a flickering effect. The term ``inflection'' was introduced in line profile studies by \cite{Hanu86}.
The inflections observed in some profiles have been attributed by \cite{Hanu86} to the profile being composed of two components, one broad and strong, and another narrow, weaker and double-peaked. In a later work by \cite{HaHu96}, the authors attribute the inflections to the viewing angle.

\begin{enumerate}
\item[]Several questions arise: 
\item Could triple-peaked profiles be thought of as an evolved state of flat-topped profiles?
Do inflection points and wiggles in flat-topped profiles have anything in common?
\item[Yes to both questions.]The evolution with the orbital phase of the profile shape can be pictured as double-peaked$\rightarrow$flat-topped$\rightarrow$triple-peaked$\rightarrow$flat-topped$\rightarrow$double-peaked (see for example the second panel of figure 3 in \citealt{SORB07}).
Furthermore, \cite{SORB07} report that triple-peaked profiles occur at phases of V$>$R to R$>$V transitions. This remark strengthens the notion of a relation between triple-peaked and flat-topped emission-line profiles.
\item Could triple-peaked profiles be seen as profiles with six rather pronounced inflection points (resulting in three emission maxima and two emission minima between them)?
\item[Yes,]according to the mathematical definition of inflections. In this case, it is logically deduced that the minima in triple-peaked profiles can be caused either by lack of emitters or by excess of absorbers at a given velocity.
The correlation of inflection points with peaks from optically thin lines hints at the latter (see, for example, figures 26-27 in \citealt{HaHu96}: the change of curvature in the $\Ha$ profiles coincides with maxima of \ion{Fe}{ii} emission).
\end{enumerate}

From all simulations presented here, quite a few demonstrate flat-topped profiles and inflections in certain phase ranges. Thus, within a certain viewing angle range that depends on system parameters, such profile shapes should in general appear in binary systems, even if no other variability source is present. Thus, also triple-peaked profiles do not have to be limited to Be
shell stars, as was suggested by \cite{SRCl09}.

Despite the nowadays commonly accepted concept that triple-peaked profiles are associated with disc warping in misaligned binaries \cite[see discussion by][]{MoNO13}, it has been shown here that such shapes can occur also in coplanar binaries.
This is not a complete contradiction: The idea rests on warped discs that can be caused by the phase-dependent vertical force from the companion in misaligned systems. Warping essentially leads to a vertical re-distribution of the disc matter in the radial direction from the non-warped to the warped region.
But this also happens in all spiralled disc structures and in coplanar binary systems, as well: The spiral arms produce a density structure that changes non-monotonically in the radial direction (see for example figure~4 in \citetalias{PCO15}).

The reader is reminded that profiles as the ones shown in \autoref{f:obsdsq1} should not be exclusively associated with binaries with mass ratio $\qr\simeq1$. The system simulated to produce those profiles has specific stellar, disc and orbital characteristics. Changing the value of each one of them causes modifications or even disappearance of certain observational features.
This of course applies to all results shown earlier in this text (for example, flat-topped profiles should not be necessarily attributed to low viscosities, but with more pronounced spiral arms).

\subsection{V/R curves for different emission lines}\label{s:pdiff}
Figures \ref{f:obsHao}k-l show the emission line profiles and the V/R ratios in $\Ha$, $\Hb$ and $\Bg$ for the low-viscosity system (\#42) at $i=90\degr$. The relative flux is smaller for $\Hb$ and $\Bg$, and so is the variability amplitude of the V/R ratio.
In higher viscosities and lower inclinations there is hardly any emission and V/R variability. This might be an indication that $\Hb$ and $\Bg$ originate from a smaller region of the inner disc, and are thus less perturbed by the secondary.

Phase lags between V/R curves have been reported for various pairs of emission lines (many relevant observations are mentioned in \citealt{Baad85}).
Their existence is due to the helical disc structure \citep{Okaz91}. For the pair $\Ha-\Bg$ they can be attributed to the smaller emission region of $\Bg$, as \ion{H}{i} IR lines are optically thinner \citep{WiKo07}.
The V/R cycles of $\Hb$, $\Hc$ and $\Hd$ reported by \citet{Dods36} for 25\,Ori exhibit phase differences. In fact, \citeauthor{Dods36} (her figure 4) reports $\Hb$, $\Hc$ and $\Hd$ V/R cycles of \mbox{$\simeq5.5$ yr} for 25\,Ori, close to the $\Porb$ value roughly estimated in \S\ref{s:ftp}.
The strong variability of the strength of emission lines documented in the said figure makes this a moot point: The inferred variable mass injection rate prevents QS to be reached so that there is no phase lacking.

The phase shift between different emission lines is a result of both the size of the emitting region and the azimuthal structure of the disc. If an emission line is only emitted close to the star, its emitting region will be smaller. The size of the emitting region is important for the emission peak height, but not its variability. The spiral structure of the disc defines its azimuthal modulation and thus the orbital modulation of observables.
The spiral arms are essentially accompanied by radial modulation of the density along a constant azimuthal angle, and result in phase lags of the V/R ratio of different emission lines.

\section{Conclusions}\label{s:end}
Perturbed line profiles that deviate from the standard double-peaked profiles of axisymmetric discs have in the past been suggested as related to binary tidal effects.
The present work demonstrated that this is a very plausible scenario: binary-induced disc perturbations result in V/R variations of spectral lines.
Under special circumstances, even a single emission-line profile can provide hints at the binary nature or not of a Be star. Time series of profiles may allow to constrain the physical properties of binaries.

\begin{itemize}
\item[]Binarity causes a two-armed spiral structure in the Be disc. Depending on system parameter values, this structure is accompanied by certain observational features. The main conclusions regarding those features are the following:
\item Disc truncation results in reduction of the emission peak heights because they form at low orbital velocities, which is the region cut by the companion.
\item Two main characteristics seem to prove binarity: If the star is seen equator-on, there will be relatively strong V/R variability. Otherwise, flat-topped profiles are indicators of the existence of a binary companion. None of those rules can be applied if the disc viscosity is large.
\item Flat-topped $\Ha$ profiles appear at intermediate viewing angles. They seem to be closely associated with profile wiggles and triple-peaked profiles, and they occur during blueward and redward transitions of double-peaked profiles.
\item Along the orbital cycle the V/R ratio exhibits two maxima at which the double-peaked emission profiles are maximally blue-shifted. The generally unequal maxima of the V/R ratio are associated with one of the spiral arms being more pronounced than the other, contributing more strongly to the total emission.
\end{itemize}

Binarity is only one of the possible means to produce variability in Be stars. Even if binarity is indeed the cause of V/R variations, it often may not be strong enough, so that the binary signature becomes veiled by other variability sources. It is still unknown what is the exact origin of mass ejection from the stellar surface. The most probable scenario is that mass is ejected due to instabilities caused by non-radial pulsations (NRP; \citealt{RiBa03}), in a probably non-continuous and non-isotropical manner.
The NRP frequencies are connected to the short-term variability of Be stars \citep{Brite1}, probably because of the non-axisymmetric ejections of mass that they cause. The rapid rotation of Be stars assists the mass particles with escaping the photosphere, which makes the disc start being built up. This causes azimuthal and temporal variability due to wave propagation.
Some memory of this variability may be preserved by the mass ejection mechanism and so induce time dependencies. The most important variabilities are long-term modulations, probably also caused by NRPs, of the mass loss rate.

As soon as the disc has reached some adequate extent and mass, it starts to oscillate. One-armed density waves due to disc oscillations (which might as well be triggered by NRPs) or/and two-armed spirals due to the gravitational effect from a potential companion are ways to maintain (or complicate) the variation scheme.
The two-armed spiral structure can be found only in binary systems, but this only adds to the effects from NRPs, stochastic and localised ejections, one-armed global density waves, all of which can exist in both single and binary Be stars.

\paragraph*{Acknowledgements.}
The authors thank the anonymous referee for prompting the extension of this study, which improved greatly the quality of the manuscript.
This work made use of the computing facilities of the Laboratory of Astroinformatics (IAG/USP, NAT/Unicsul; brazilian agency FAPESP, grant 2009/54006-4). The authors also thank for the much-needed access to the computer cluster of the Group of Applied Geophysics (ON, Brazil). DP acknowledges FAPESP (2013/16801-2) and CNPq (\mbox{MCTIC}, Brazil; 300235/2017-8).
DMF acknowledges FAPESP \mbox{(2016/16844-1)}. ACC acknowledges CNPq (307594/2015-7) and FAPESP (2015/17967-7).

\bibliography{biblio}
\end{document}